# Functional Zr, α-Ta, and (α-β) Ta thick coatings obtained by original arc-evaporator.

K. Pavlov [a], P. Vuoristo [b], A. Savin [c], H. Koivuluoto [b],
K. Gorchakov [a].

[a] - *Kraftonweg Oy, Ruuvikatu 6, 48770, Kotka, Finland*
[b] -*FASM, Tampere University of Technology, Department of Materials Science, PO Box 527, FI-33101 ,Tampere, Finland*
[c] - *Low Temperature Laboratory, Department of Applied Physics, Aalto University, PO Box 15100, FI-00076 AALTO, Finland*

## Abstract.

Novel plasma-clustered deposition technique, called PVD Droplets, as well as its main principles of operation were described. New types of obtained thick, over 20 μm, functional coatings basing on Ta and Zr have been presented. Results of research in physic-chemical properties of the coating materials are listed including preliminary studies of superconductive properties. Comparative evaluation between of obtained coatings and sample CVD coating was made.

## 1. Introduction

Corrosion protective metallic coatings found its wide application in industrial production of machinery and equipment. Nowadays the coatings in thickness of above 20μm are obtained by various methods [1]: electrochemical deposition; chemical solution deposition; chemical vapor deposition CVD; thermo-spaying; and other ones.

Such methods require a lot of raw materials and energy followed by high costs for recycling of disposal materials. In most cases, the coating process itself as well as chemicals involved could be hazardous for human being and therefore are highly undesirable in terms of environment protection.

The original and novel deposition technique for corrosion and wear resistant thick protective coatings based on Zr, Ta, Cu has been developed by Finnish company Kraftonweg Oy in cooperation with Tampere University of Technology. The technique requires reasonable amounts of raw materials and energy featuring high added value for final products. It also does not require any harmful chemicals and is certainly friendly to environment. The technology is based on the principle of metal evaporation induced by cathode spot, known as ARC–PVD evaporation method [2]. The process of coating takes place in high vacuum of below $P=1\times10^{-2}$ Pa. Arc evaporation of single or even of several metals, results in formation of plasma-clustered flows of various chemical elements, in other words, the plasma flows consisting of material elements which after mixing up form uniform flow to deposit onto substrate.

Condensation of plasma-clustered flow can be in conditions of extremely rapid dissipation of thermal energy being the subject to required properties of obtaining coating material. As a result the coating material can be formed in amorphous, homogeneous and extra fine-grained structures as well as in normal polycrystalline one. It is noteworthy that non-stoichiometric materials could be also attained by this method. The deposition rate was reached up to 1 micron per minute.

## 2. Coating technique

Principal distinction of used original method from most 'industrial' PVD ones is namely the deposition of unseparated plasma-clustered flow straightly at specimens, with three flow components could be contingently recognized: plasma, vapor and micro droplets, whilst their percentage depends on both cathode material and deposition parameters.

Preparatory measures for samples/ machinery components require simple solvent treatment removing oily residues and further subjecting to sandblasting and compressed air blowing.

For implementing of plasma-clustered deposition the original outspread evaporation system providing elongate sectoral material flow has been developed. Its schematic design is presented on Pic.1.

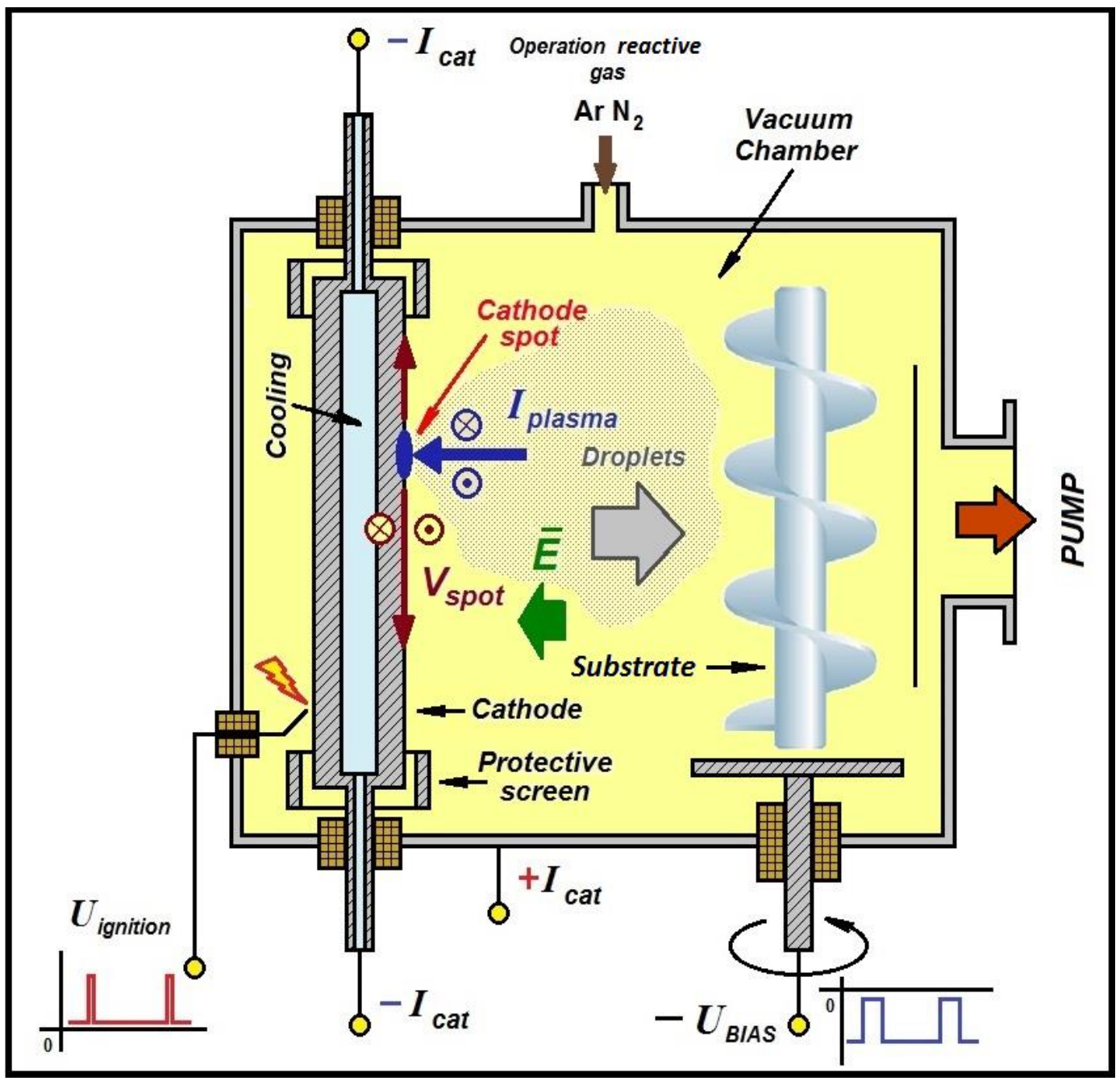

Pic.1. Design of evaporation system.

Once the specimens or machinery components were fixed inside vacuum chamber and operational vacuum has been reached the process of heating and surface activation starts. Surface activation is performed by high energy ions of argon beams resulting in final deletion of contaminations and also heating up the samples to desired temperature.

Then the namely deposition process is activated involving outspread arc-evaporator made of the material chosen as subject to required coating material and/or its structure. High voltage ignition initiates several cathode spots enabling evaporation of the cathode material. The cathode spots are forced to move up and down along cathode generatrix so that they are faced against samples. The screens placed at top and bottom of the cathode serve to keep the spots in operation area preventing them from creeping down to non-functional locations. Velocity of cathode spots relocation reaches 50 meter per second while the samples fixed in vacuum chamber planetary rotated, so both these factors provide ultimately uniform deposition of evaporated materials.

In case the gases are supplied the plasma-chemical reactions lead respective ions to be imbedded in coating material, again subject to coating material and structure required.

Thus, homogeneous plasma clustered flow together with (or without) gases supplied provide much flexibility in structures and composition of coating materials even covering attainable solid, gradient and layered materials.

## 3. Experimental

The properties of obtained specimens have been studied by FE-SEM , X-ray structural analysis, hardness measurements and corrosion resistance tests. Micro structure has been investigated by optical microscope MMP-4 and Scanning Electron Microscope Zeiss ULTRAplus. The micro hardness at cross section tests have been performed with MMT-X7 Vickers microindenter, Matsuzawa, Akita, with applying loads of 100g and 25g.

Structural and phase compositions of the coatings have been determined by X-ray analysis on diffractometer XRD, Empyrean, PAN Analytical, Cu-Kα radiation.

### 3.1. Samples under research

Four specimens were subjected to experimental research:

- The reference one, pure tantalum coating Ta(Ref) as a model to compare.

The sample represented the 30mm diameter washer made of stainless steel AISI 316 with Ta coating produced by CVD method. This type of coating has rather wide applications in machine building industry and also for machinery components.

- Two-layered (α-β)TaN coating deposited on 8 x 30 x 60mm sheet of stainless steel AISI 316. The deposition was performed by PVD Droplets technology.
- Single-layered Ta(Ti) coating deposited by PVD Droplets technology on a plate made of technically pure Titanium plate in sizes of 8 x 30 x 60 mm.
- Single-layered Zr coating deposited by PVD Droplets technology on a plate cut of sheet stainless steel AISI 316 having sizes of 8 x 30 x 60mm

### 3.2. **Structure and thickness of the coatings**

The below photomicrographs of Ta(Ref) coating cross section present: structure and overall thickness of the coating Pic.2(a); structure of inside of coating material at higher magnification Pic.2(b); quality of adhesive contact i.e. of area between coating and AISI substrate. Pic.2(c).

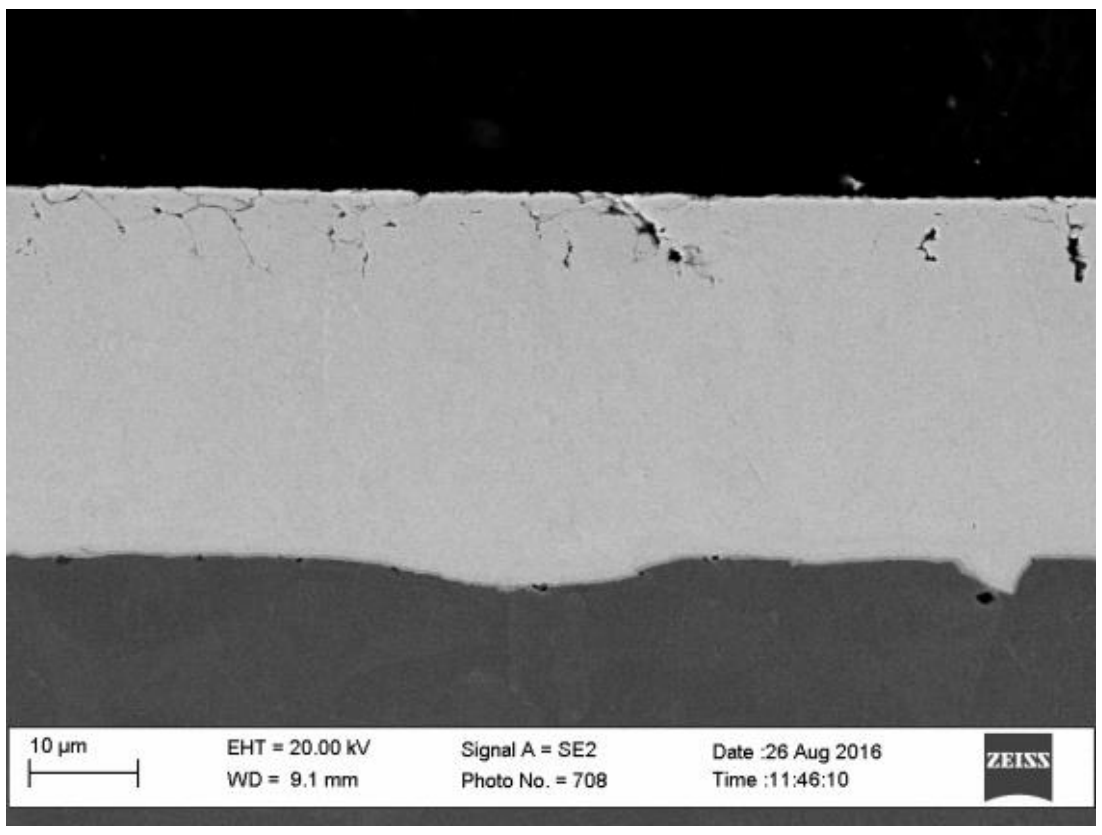

Pic.2(a). SEM Photomicrographs of Ta(Ref), overall view.

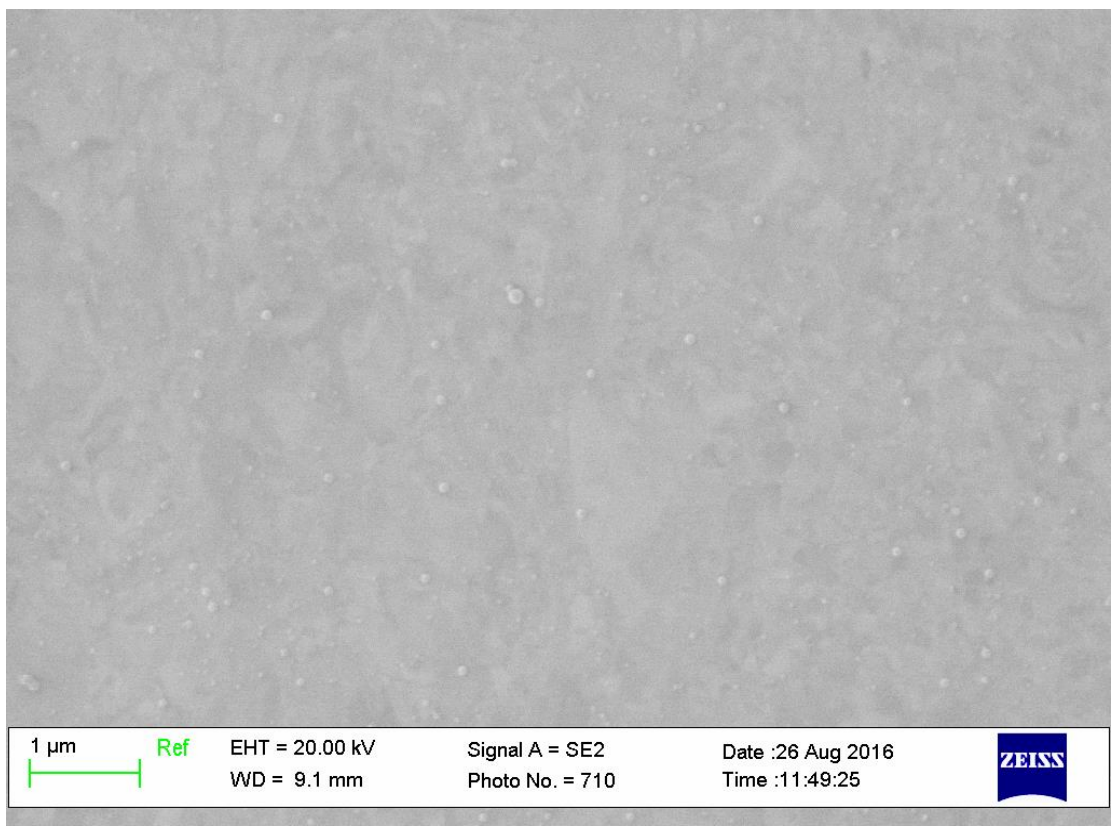

Pic.2(b). SEM Photomicrographs of inside of Ta(Ref).

The defective subsurface layer spreading into material is as deep as about 10µm as observed on Pic.2(a). The defect-free layer region is visible of about 25µm thickness Pic.2(b) being fine grained and having homogenous crystal structure. At the adhesion area one can see the additional, presumably adhesive layer Pic.2(c) providing reliable adhesion between the coating and substrate. Minor number of pores between the coating and substrate evidences of good preparation of the surface before coating process.

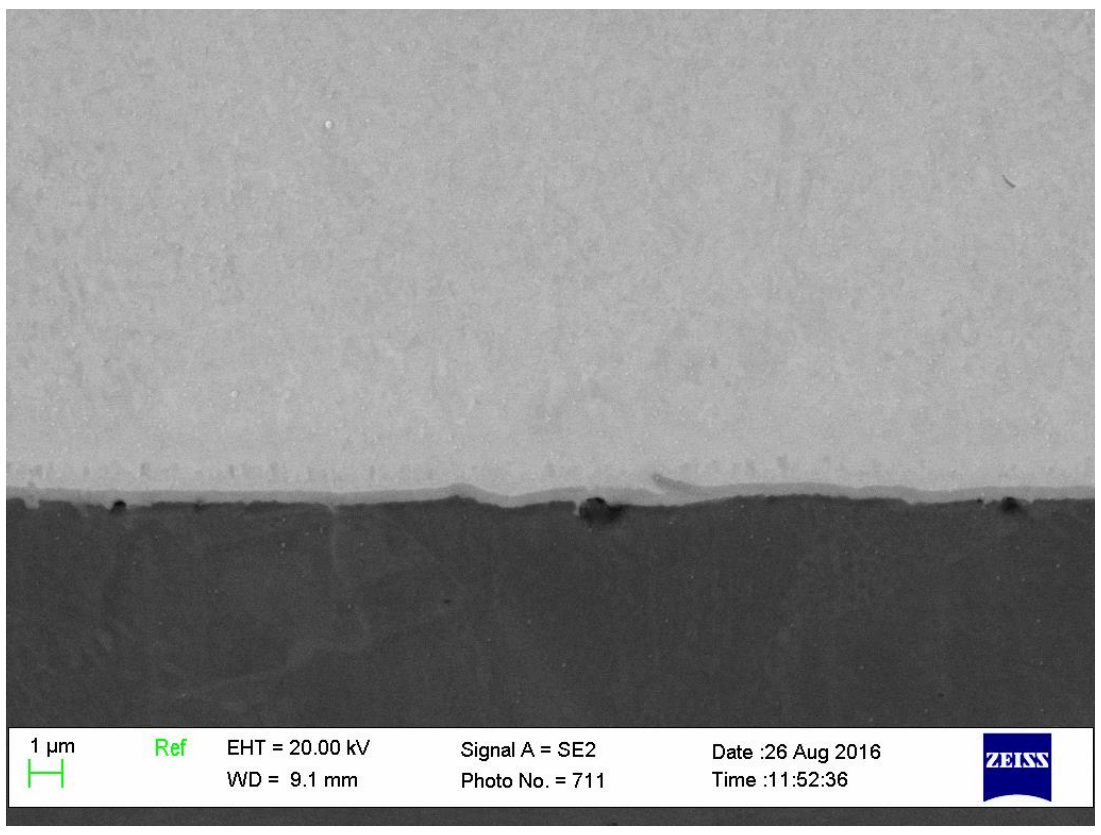

Pic.2(c). SEM Photomicrographs of adhesion area of Ta(Ref).

Pic.3 represents cross-sections photomicrographs of (α-β)$TaN_{(0,1)}$ coating in thickness of about 100 µm and having two different sublayers.

The bottom one adjoins AISI substrate and is about 40µm in thickness constituting of α-$TaN_{(0.1)}$ with body-centered cubic structure (bcc) [3].

The upper sublayer of about 60μm thickness is hard and exhibiting relatively good elasticity. This sublayer of β-$TaN_{(0.1)}$, non-stoichiometric tantalum nitride, is featuring by pseudo-cubic distorted crystal lattice [4] [5].

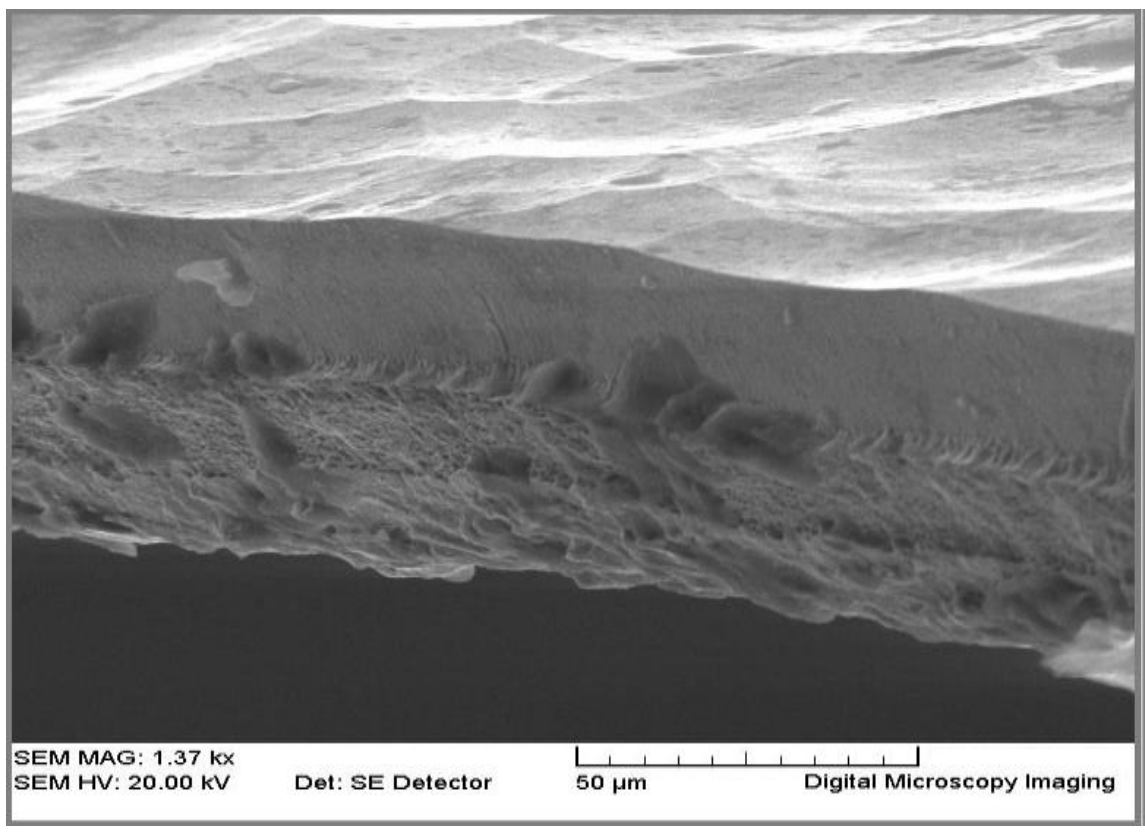

Pic.3(a). Cleavage cross-section of (α-β)$TaN_{(0,1)}$ view 1

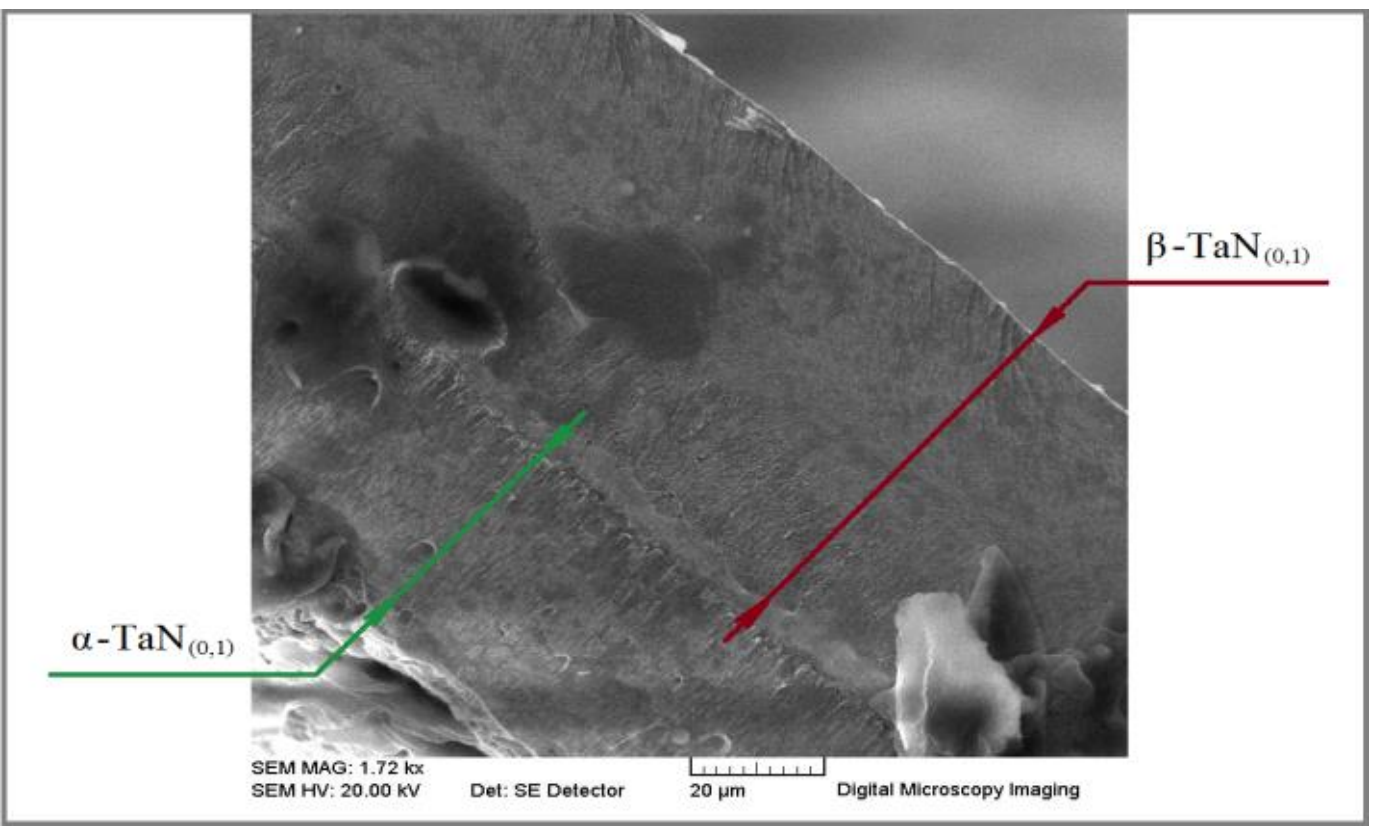

Pic.3(b). Cleavage cross-section of (α-β)$TaN_{(0,1)}$ ,view 2

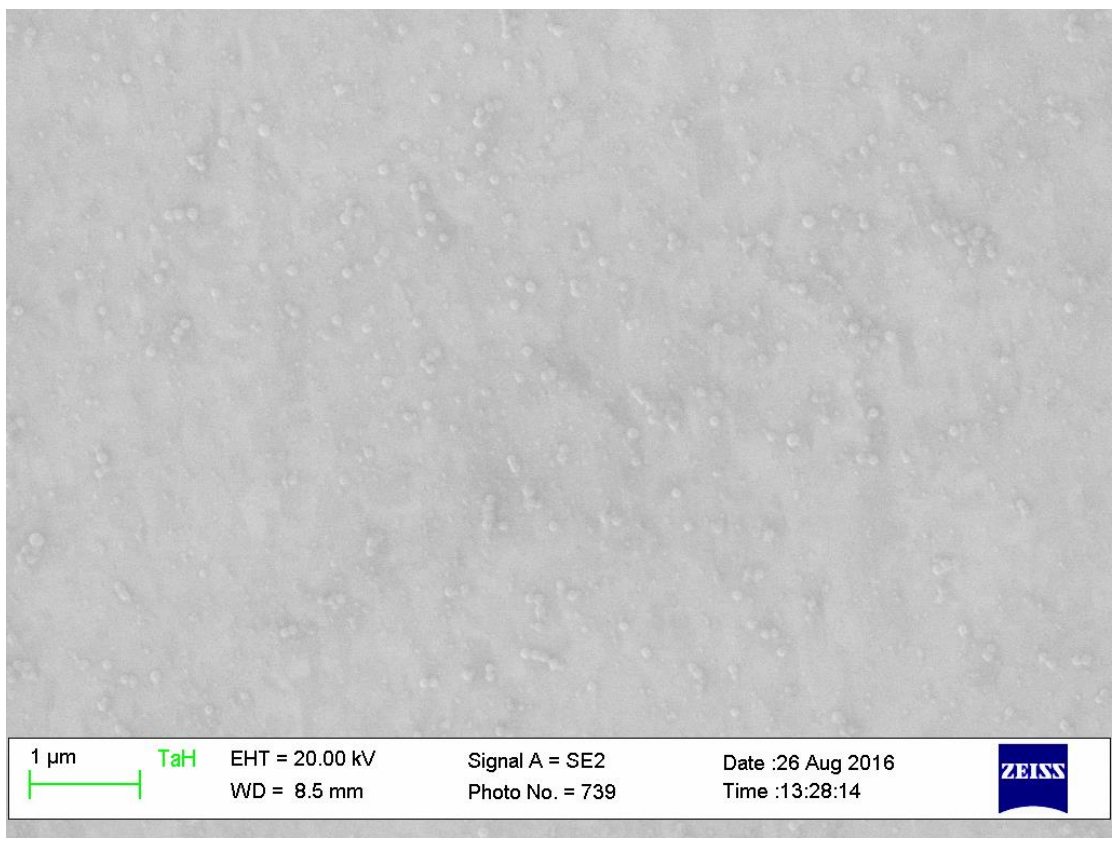

Pic.3(c). Polished cross-section of (α-β)TaN(0,1) , view 1

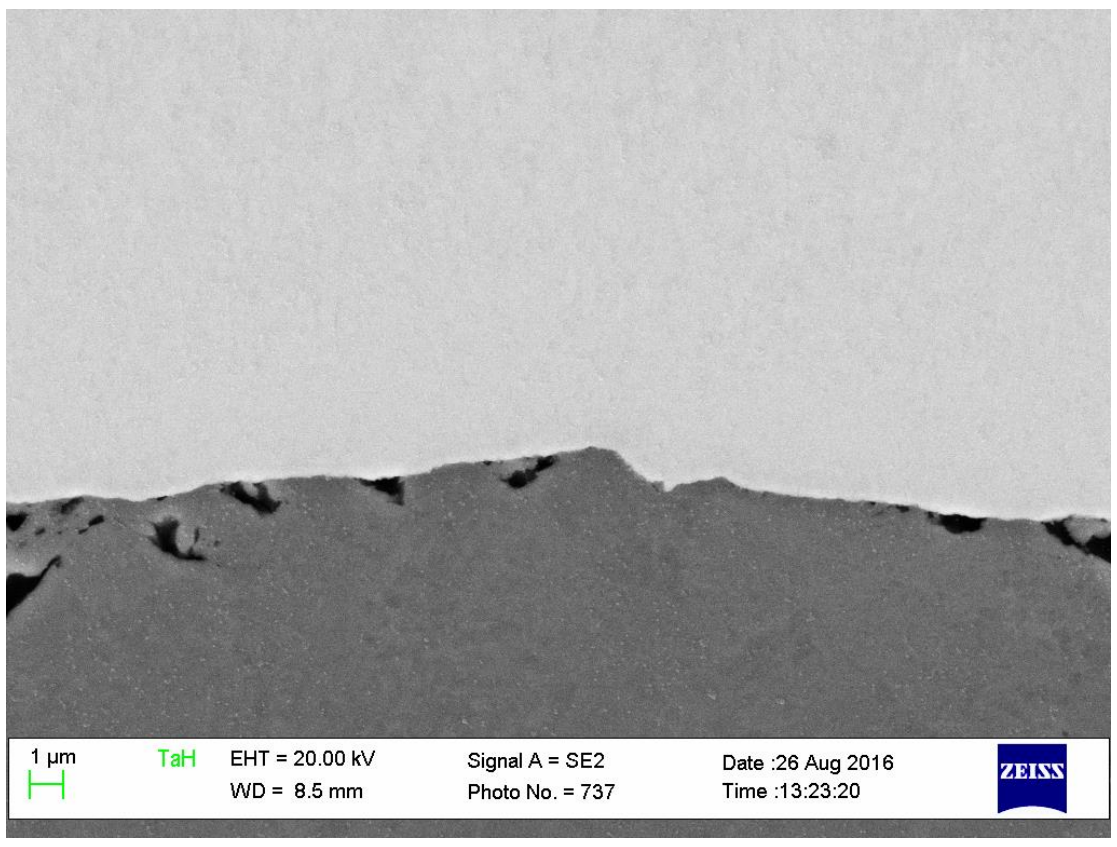

Pic.3(d). Polished cross-section of (α-β)TaN(0,1), view 2

Pic.3(a) and Pic.3(b) exhibit different structures of sublayers on cleavage made through simple bending of detached coating until fracture. The photomicrograph Pic.3(c) of polished cross section shows rather fine grained homogenous crystalline structure, the structure looking very similar to Ta(Ref) Pic.2(b). The area adjoining to substrate displays no any intermediary adhesion layer and also having a little more concentration of pores and cavities than Ta(Ref) coating.

Single-layered Ta(Ti) coating deposited onto Titanium lamella of 99.5% purity and its cross section is presented below on Pic.4(a), Pic.4(b) and Pic.4(c).

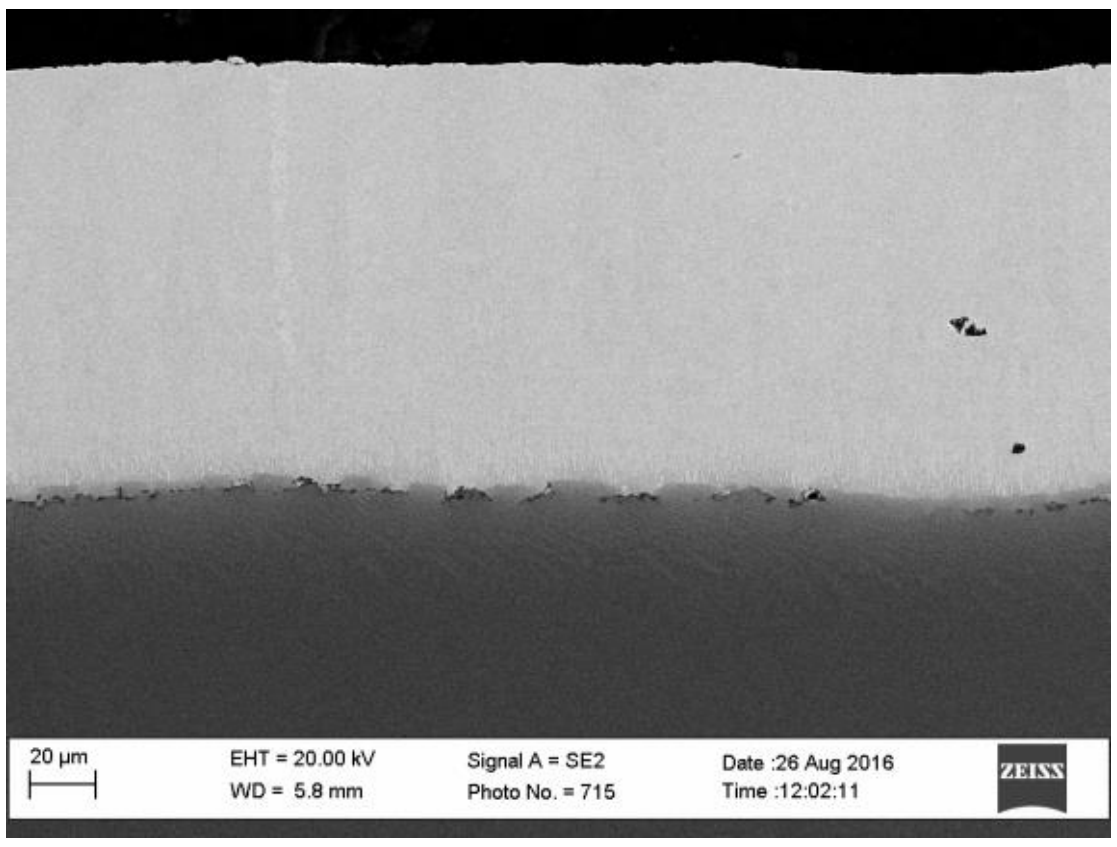

Pic. 4(a). Cross-section of Ta(Ti)

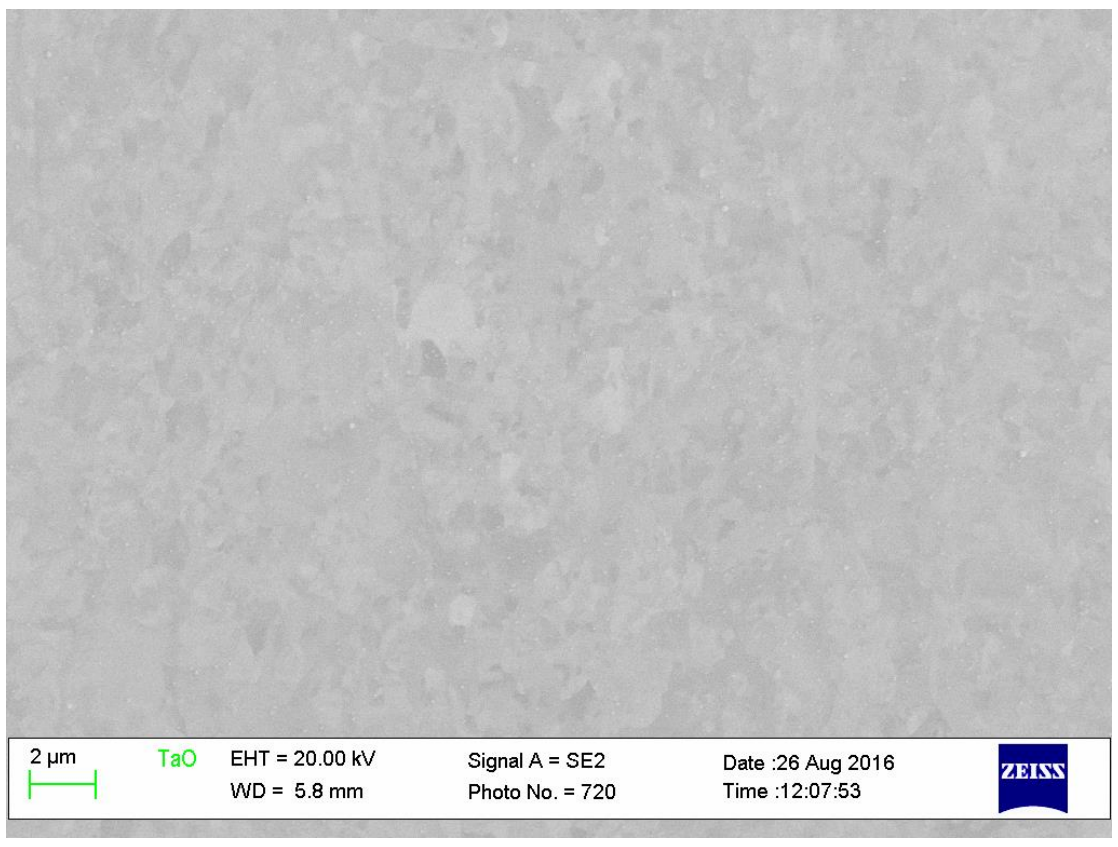

Pic.4(b). Cross-section of Ta(Ti), higher magnification

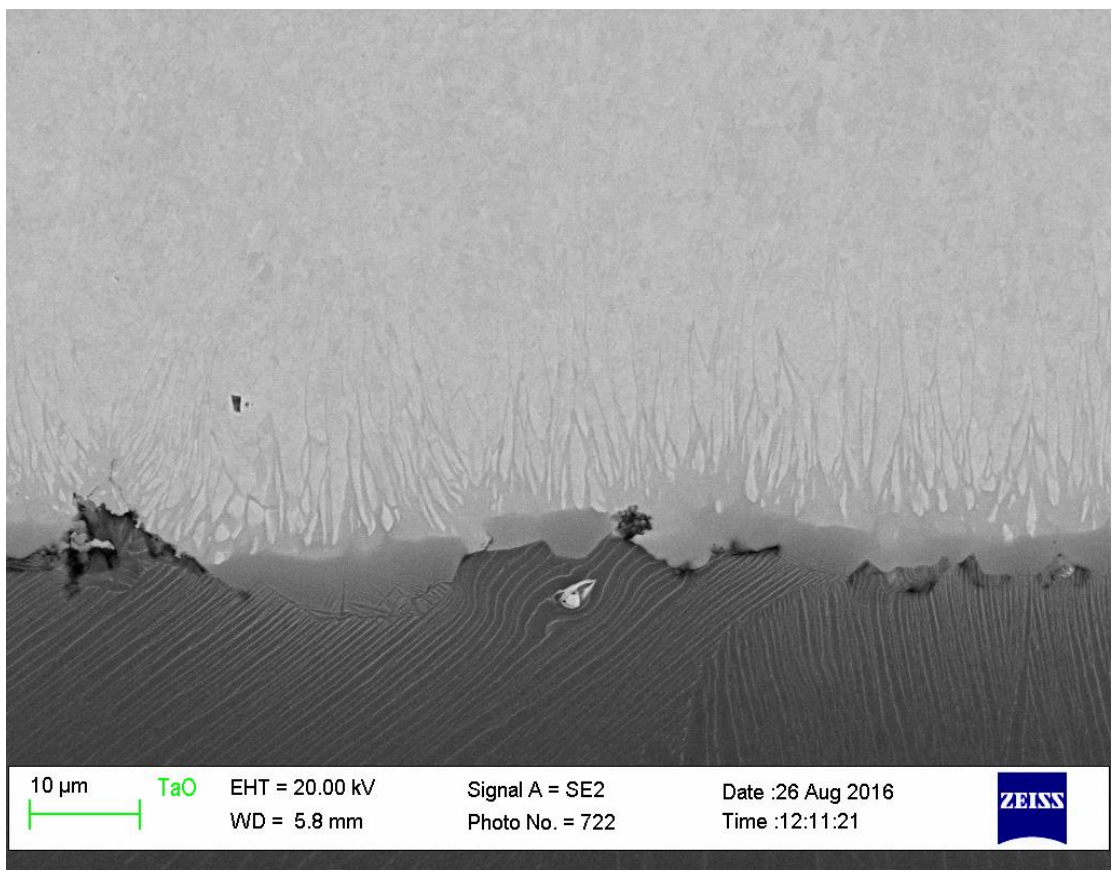

Pic. 4c. Cross-section of Ta(Ti), bottom area.

Solid, single-layer, homogeneous, and 120μm thick coating consists of (bcc) α-Ta and apparently reveals no defects like pores or caverns in the main thickness area of the coating. The internal structure visible on the coating cross section is analogous to the above described samples at the same magnification (Pic.4b). The most interesting is the area visible on (Pic.4c) where α-Ta coating adjoins titanium substrate. As has been specified by X-ray analysis this descending type structure has been formed under influence of oxygen, nitrogen and other chemical compounds released from inside of Titanium substrate subsurface at high deposition temperatures in vacuum [6]. Oxides, nitrides and other compounds near substrate result in very low mechanical strength and therefore tend to possible delamination of the coating when subjected to hard industry applications.

The following Pic.5(a), (b) and (c) exhibit single layer of solid zirconium coating deposited onto lamella on stainless steel Pic.5(a). The thickness of the coating is about 200μm.

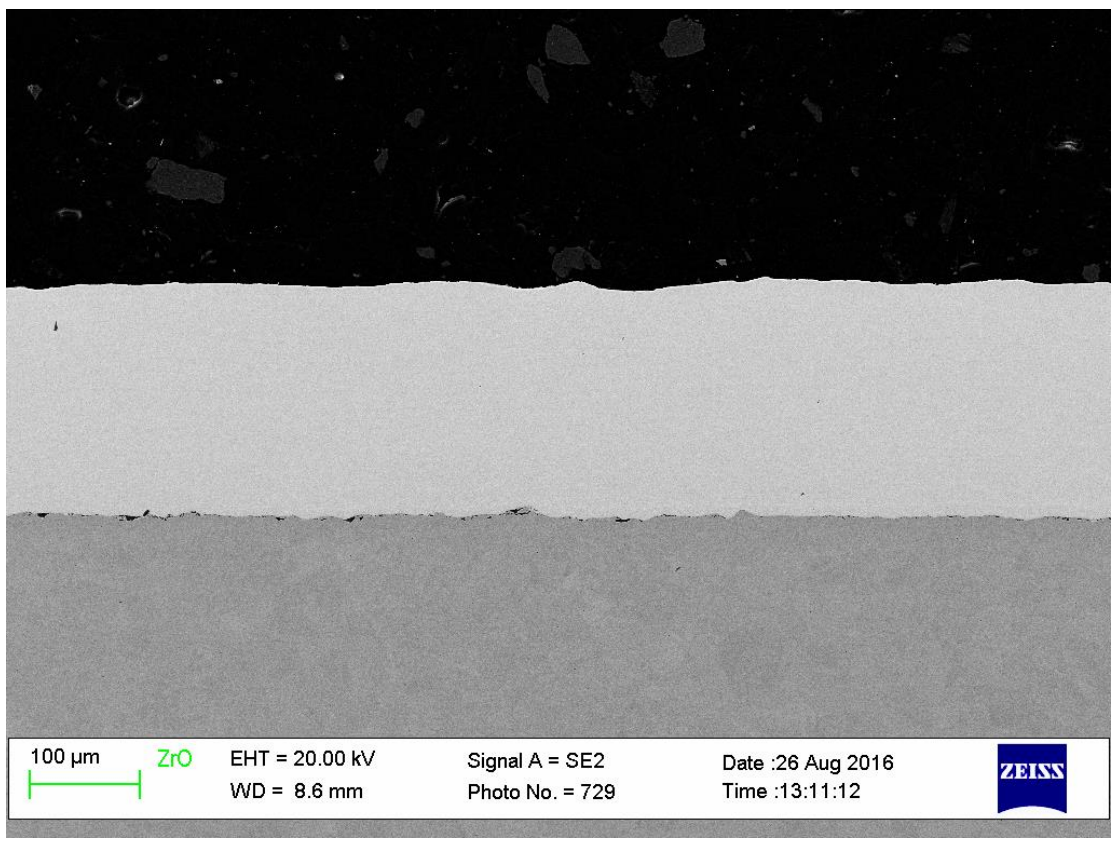

Pic. 5(a). Cross-section of Zr coating

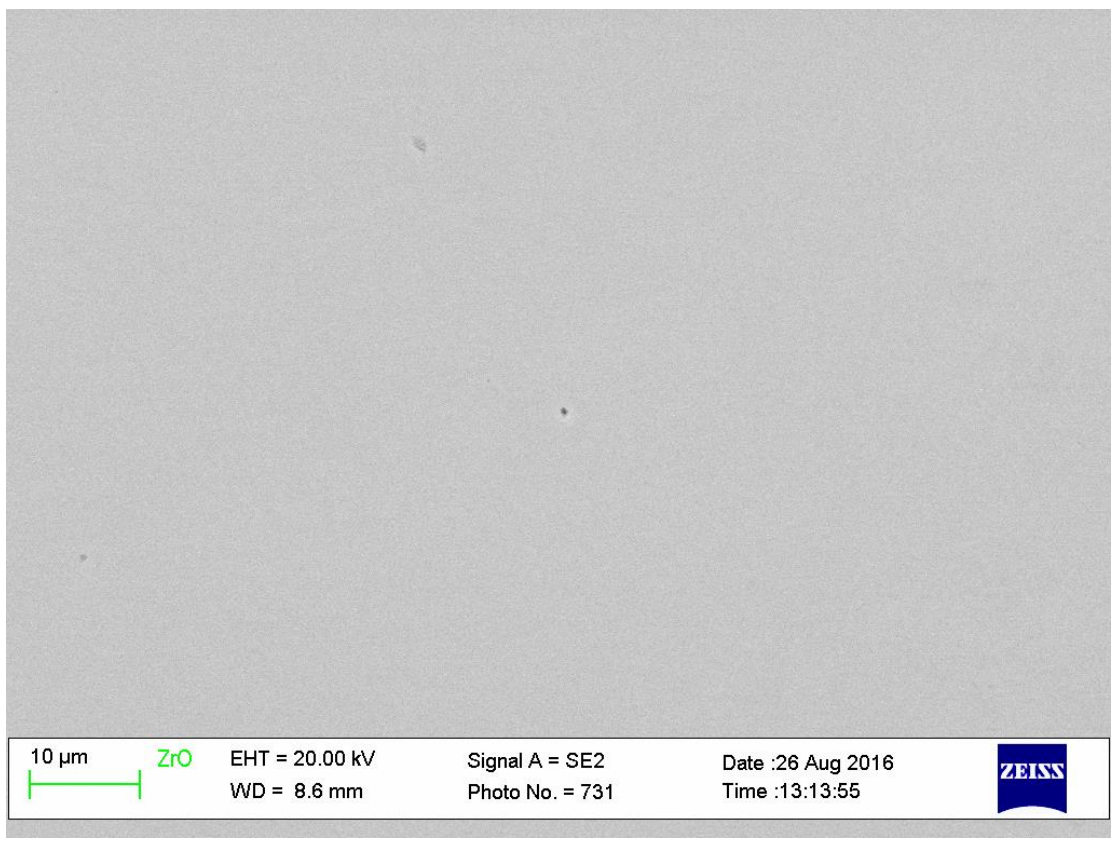

Pic.5(b). Cross-section of Ta(Ti), higher magnification

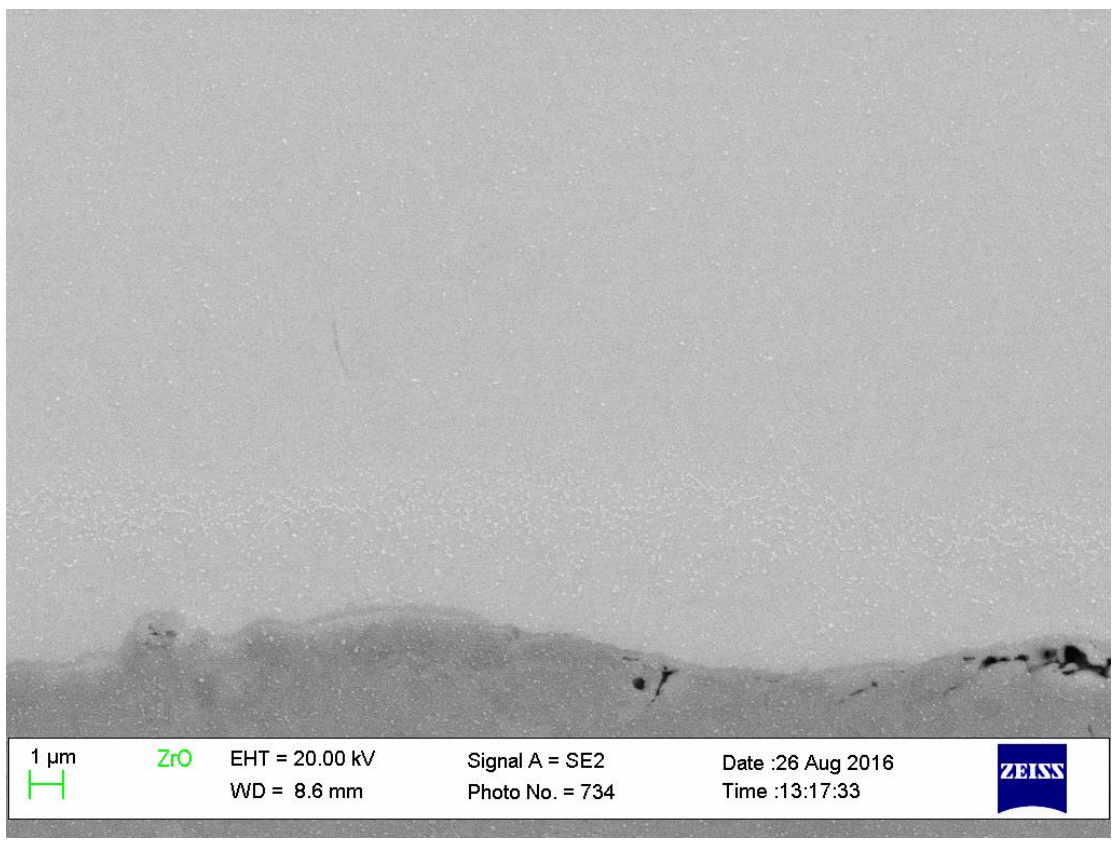

Pic.5(c). Cross-section of Zr coating.

The structure at the middle of coating shown on Pic.5(b) features extra fine grained, sub-crystalline structure practically amorphous one. No grain boundaries are noticeable. The material is extremely homogeneous, practically defects free. The adhesive zone is presented on Pic.5(c) with no visual adhesion layer. Visual analysis reveals a little less defects concentration in adhesion zone to compare with the same area of $TaN_{(0.1)}$ Pic.3(в).

### 3.3. Analysis of crystal lattice by X-ray diffraction method (XRD)

The phase composition of four PVD and one CVD thin coatings was assessed by X-Ray Diffractometry. Experimental conditions include 2θ range 20-80°, step size 0.0260°, irradiated length 20 mm, and PANAnalytical PIXcel 3D detector. Phase identification was performed using the PANAlytical X'Pert High Score Plus software using the ICDD JCPDF-2 database (International Centre for Diffraction Data, Newtown Square, PA, USA).

The samples were labelled as 1-TaN, 2-Ta(Ti), 3-Zr, 4-Ta(Ref) and 5-Chipped film TaN coating. Sample 5 is a free standing coating and was analyzed both on top and on bottom (Ta-top and Ta-bottom). Pic.6 presents the XRD pattern of the superimposed curves for samples Ta(Ti), TaN and Ta(Ref.) CVD.

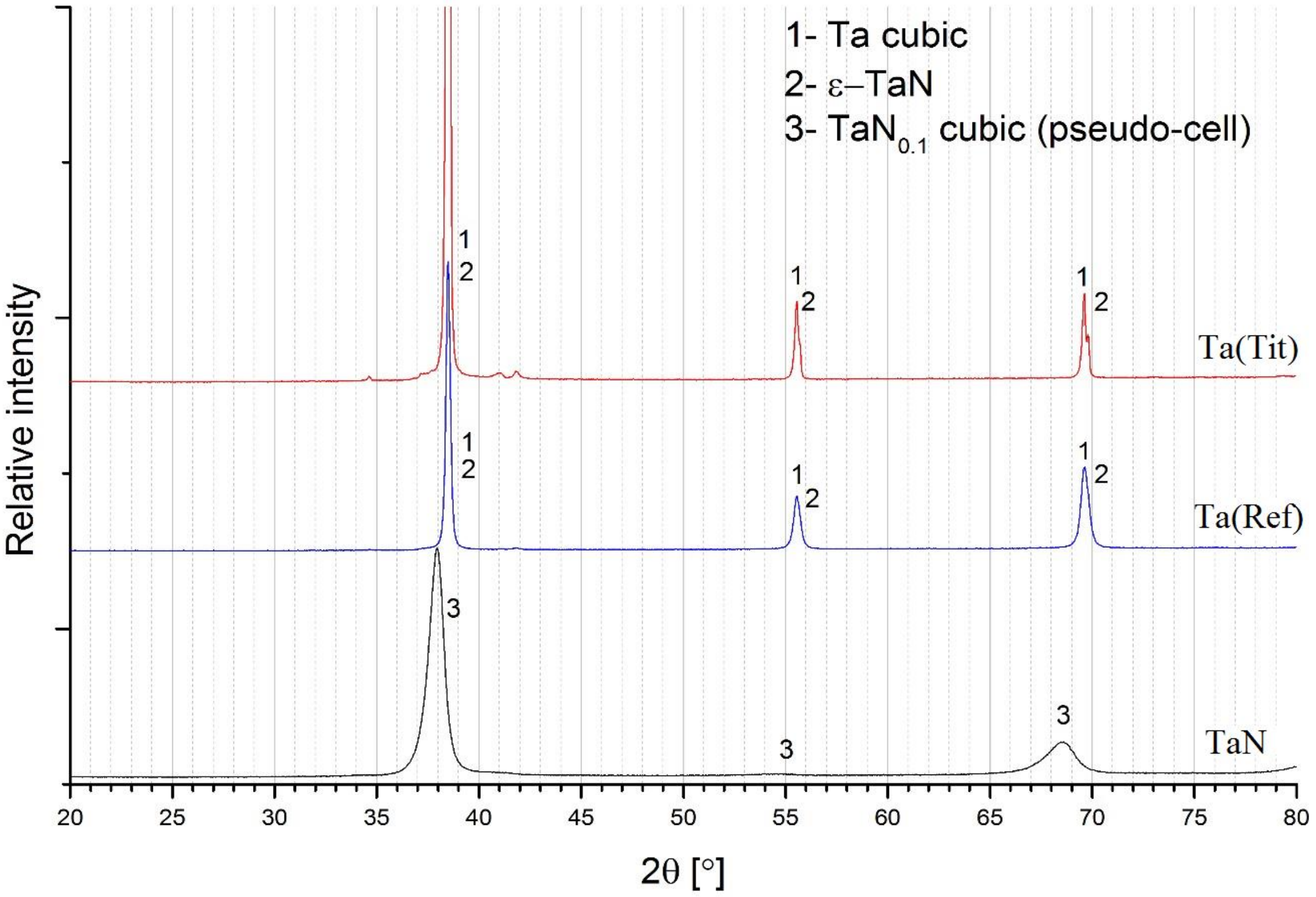

Pic. 6. XRD patterns overlays of samples TaN, Ta(Ti) and ref.

The main phase in Ta(Ti) and Ref samples is Ta cubic (ICDD ref. no. 00-004-0788) while sample of TaN seems to be composed of a non-stoichiometric $TaN_{0.1}$ body centered pseudo-cell for superlattice (ICDD ref. no. 00-025-1278). This unordinary cell structure is probably due to the distortion of the bcc cell induced by N. Sample Ta(Ti) and the reference sample from Ta(Ref) presented very similar patterns including ε-TaN hexagonal (ICDD ref. no. 04-004-3000) and few other unidentified peaks. However, the patterns are slightly shifted compared
to the standard reference which may denote the presence of non-stoichiometric phases.
Moreover, the peaks of Ta cubic, Ta(Ti) and reference samples seem to be shifted toward lower diffraction angles compared to the reference ICDD data, indicating possible expansion of the lattice, presumably induced by doping elements.

Pic.7 represents the superimposition of the patterns obtained from the different analysis of the two sides of the freestanding TaN coating. The pattern obtained from the top side seems to be identical to that of sample TaN and represent a distorted Ta cubic cell. However, the peaks of Ta-top seemed shifted toward smaller cell parameters and perhaps contain a lower amount of N. The pattern of TaN is again represented in the graph for comparison.

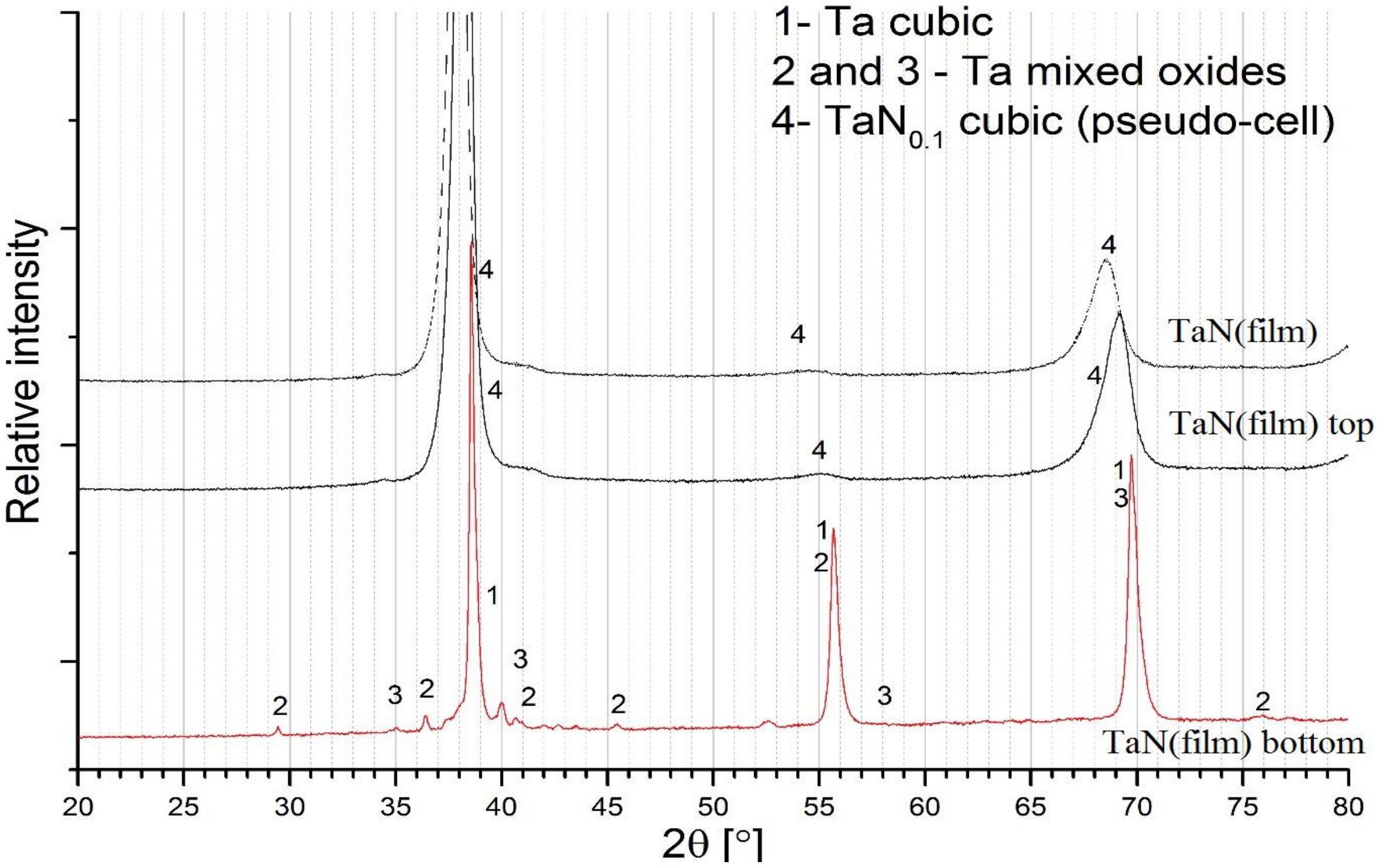

Pic. 7. XRD patterns of free standing coating sample in TaN and Ta(ref).

The pattern of the bottom side of the freestanding coating presents more peaks than the top side. Two different non-stoichiometric Ta rich oxides were indeed detected on the surface and they are also visible under SEM microscope.

The last sample is a Zr metallic coating and it differs in composition from the previous one a XRD pattern for this sample is presented in (Pic.8)

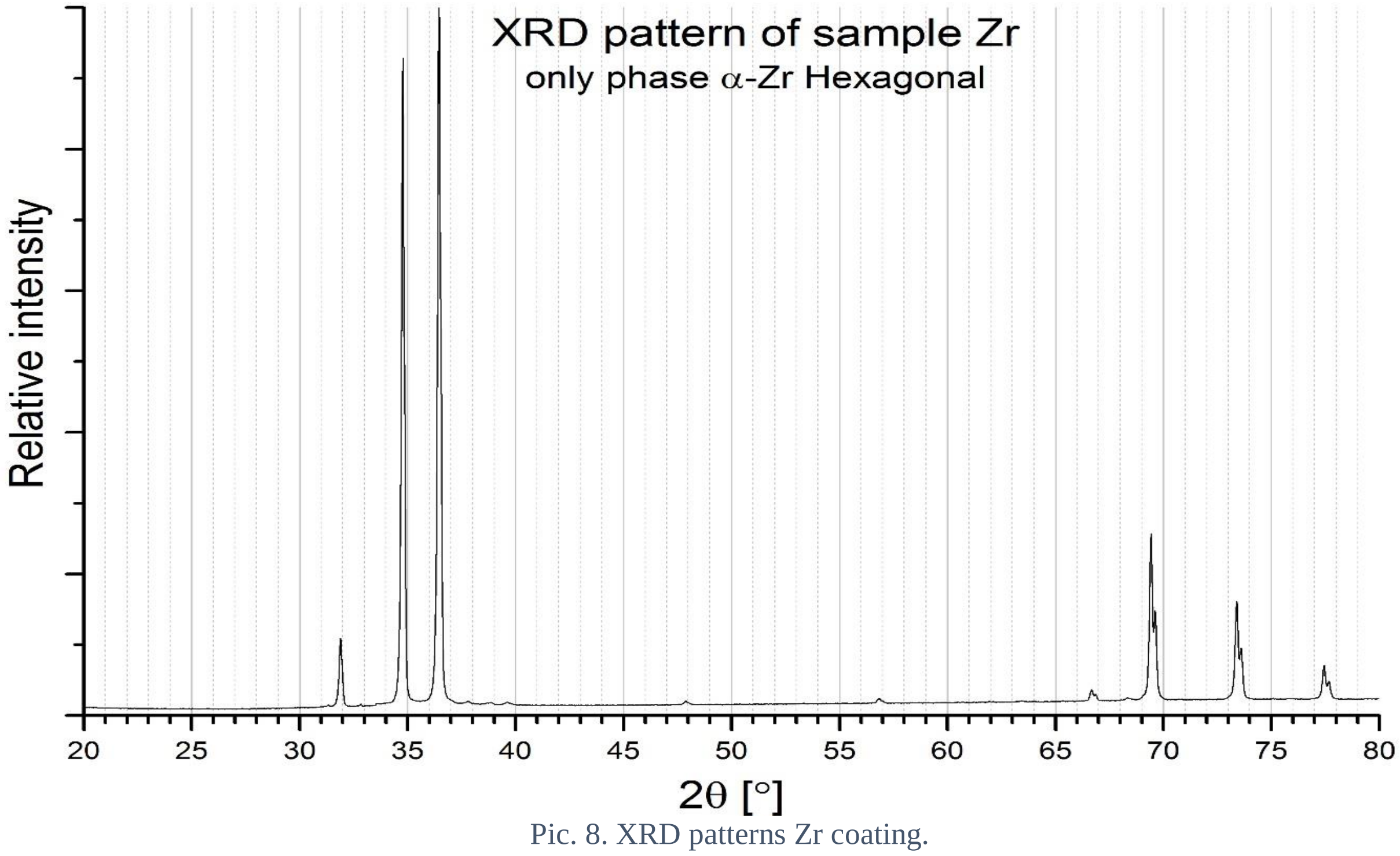

Pic. 8. XRD patterns Zr coating.

The pattern showed high purity of the coating which is solely composed of α-Zr Hexagonal (ICDD ref. no. 04-008-1477).

### 3.4. Microhardness measurements results

The micro hardness of the coatings was measured by Vickers hardness method through using a load of 100g. (HV0,1). Because of the low thickness of the reference coating of Ta(Ref) the 25g. load (HV0,025) was applied. Ten samplings were made at the middle of each layer and have been subsequently averaged afterwards. Deviation of measured values was found to be good enough - less 5%. Related results are tabulated below for each coated sample. Pic.9

| Coating composition | Ta(Ref) | $TaN_{(0.1)}$ | | Ta(Tit) | Zr |
|---|---|---|---|---|---|
| | | α - layer | β - layer | | |
| Hardness (HV) | 232 | 377 | 774 | 99 | 138 |
| Load on indenter (gram) | 25 | 100 | 100 | 100 | 100 |
| Coating thickness (µM) | 35 | 40 | 60 | 120 | 200 |

Pic.9. Micro hardness values and thickness of samples under research.

It is noteworthy that the PVD Droplets coating TaN(0.1) shows high values of hardness, significantly higher than of the reference Ta(Ref) CVD coating (300÷700 vs. 200 HV). Additionally micro hardness values of Ta deposited on titanium substrate as well as Zr upon AISI 316 one are presented.

### 3.4. Studies of corrosion behavior of the coatings by using potentiodynamic polarization measurements in two different acids.

Cyclic polarization measurements provided useful information about pitting susceptibility, passivity as well as cathodic behavior of tested samples. The measurements of both cathodic and anodic dissolving have been conducted through three-electrode cyclical technique in HCL 0,1M и $H_2SO_4$ 0,1M solutions as electrolytes.

The electrode voltage (Vvs.Ref.) applied to the samples was initially set to +1,0V and then has been stepwise reduced to -0,7V with increments of 0,5 mV/s. The value of overall current density (*j*-Current, $\mu A/cm^2$) of tested samples has been gauged at each step of voltage reduction.

Cathodic and anodic polarization curves were obtained for the samples and are presented on Pic.10.

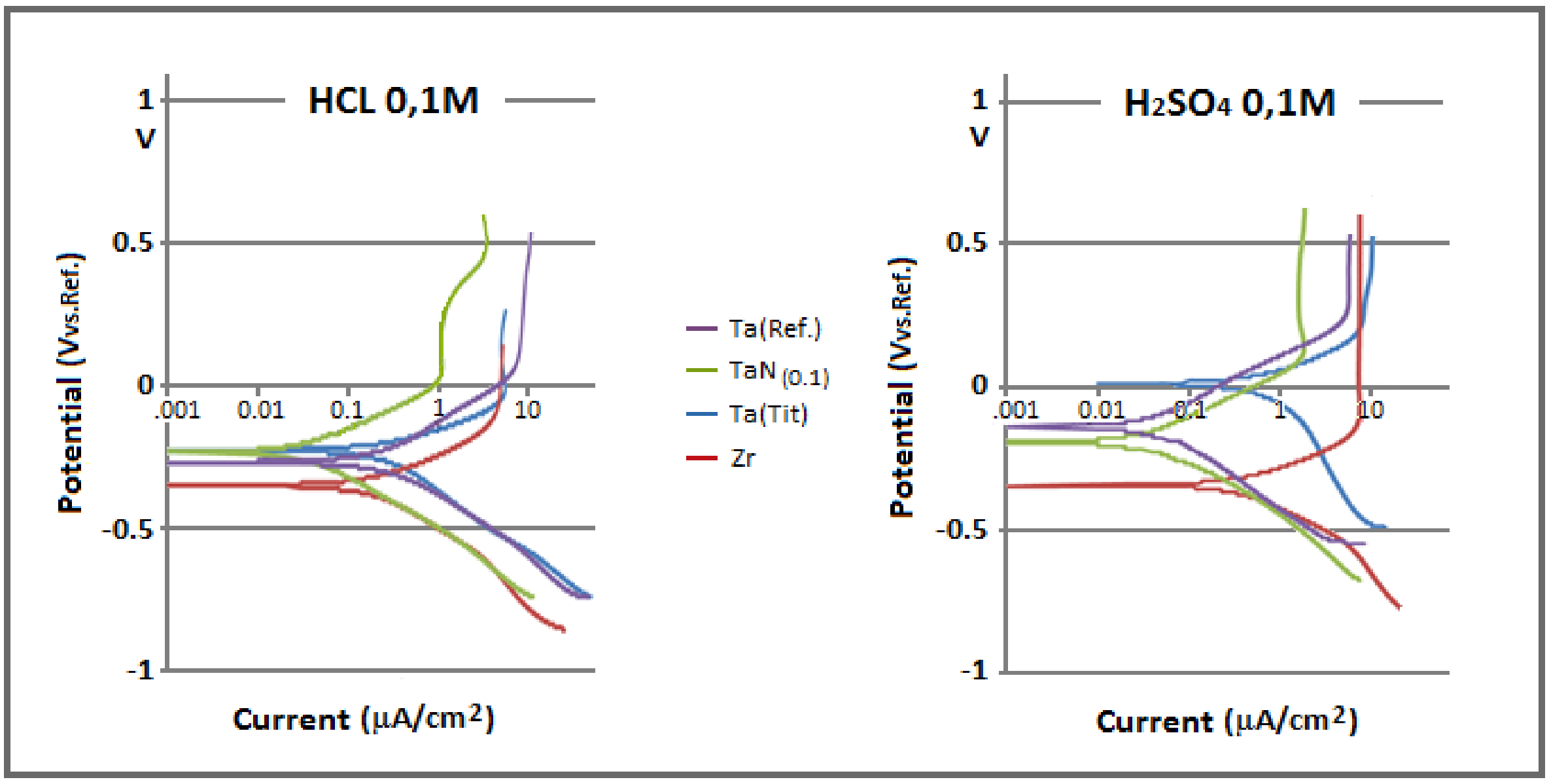

Pic.10. Cathodic and anodic polarization curves for tested samples.

Absolute value of current density ***j***-Current (µA/cm$^2$) indirectly determines chemical activity and, therefore, dissolving rate of the material submerged in electrolyte. The dissolving rate could also depend on polarity of applied potential. The higher values of ***j***-Current mean worse corrosion resistance of the material at the same values of Vvs.Ref.

Values of ***j***-Current at different Vvs.Ref. are presented in Pic.11 and Pic.12 for solutions of в HCL 0,1M and $H_2SO_4$ 0,1M respectively.

| Electrolyte HCL 0,1M | | | | | |
|---|---|---|---|---|---|
| Potential (Vvs.Ref.) | | Current (µA/cm$^2$) | | | |
| | | Ta(Ref.) | TaN(0.1) | Ta(Tit.) | Zr |
| Anodic polarization | 0,14 | 7,96 | 1,08 | 4,94 | 5,14 |
| | 0,07 | 6,95 | 1,11 | 5,07 | 4,98 |
| | 0,03 | 5,55 | 1,02 | 5,32 | 4,78 |
| Cathodic polarization | -0,03 | 2,94 | 0,61 | 5,31 | 1,78 |
| | -0,07 | 1,78 | 0,35 | 4,06 | 4,55 |
| | -0,14 | 0,8 | 0,11 | 1.32 | 3,37 |

Pic.11. *j*-Current data for HCl solution.

| Electrolyte $H_2SO_4$ 0,1M | | | | | |
|---|---|---|---|---|---|
| Potential (Vvs.Ref.) | | Current ($\mu A/cm^2$) | | | |
| | | Ta(Ref.) | TaN(0.1) | Ta(Tit.) | Zr |
| Anodic polarization | 0,5 | 5,67 | 1,74 | 10,19 | 7,41 |
| | 0,2 | 3,49 | 1,73 | 7,23 | 7,35 |
| | 0,05 | 0,37 | 1,01 | 0,69 | 7,22 |
| Cathodic polarization | -0,05 | 0,1 | 0,22 | 0,9 | 7,39 |
| | -0,2 | 0,09 | 0,01 | 2,48 | 4,02 |
| | -0,5 | 2,24 | 1,71 | 14,53 | 3,14 |

Pic.12. *j*-Current data for $H_2SO_4$ solution.

Obtained results lead to following conclusions: apparently $TaN_{(0,1)}$ coating material is the best one in terms of corrosion resistance, the next one was shown to be the reference Ta(Ref.) coating deposited by Chemical Vapor Deposition (CVD) method.
As expected both Ta(Ti) and Zr coatings showed lower corrosion resistance than of $TaN_{(0,1)}$ and Ta(Ref.) coatings. At the same time collation of Ta(Ti) and Zr coatings showed the corrosion resistance to be nearly the same for both coatings featuring rather high deviation of measured values. Thus, further research of these coatings required in order to bring more understanding.
Also worth to notice that Zr-coating revealed about 20% better corrosion resistance in hydrochloric acid than for Ta(Ref.)

### 3.5. Superconducting critical temperature of $TaN_{(0.1)}$.

X-ray diffraction analysis of crystal structure of $TaN_{(0,1)}$ coating revealed two different crystallographic phases:
1) α-$TaN_{(0.1)}$ stable (bcc) crystal phase.
2) β-$TaN_{(0.1)}$ stable pseudo-cubic phase, tense, featuring long-range atomic order.

Authors considered sensible to carry out preliminary evaluation of critical temperature ($T_c$) when superconducting phase transition takes place for in $TaN_{(0,1)}$ as far as no such results have been reported previously. It is known that Critical Temperature ($T_c$) for tantalum is 4,38 ºK while stoichiometric $Ta_2N$ has no superconducting phase transition [7]. The measurements of ($T_c$) were made by recording resistance vs. temperature using 4-point probe method.

Electrical characterization of the samples was performed in a dry 3He/4He dilution refrigerator enabling measurements at variable temperatures down to 12 mK. Samples under investigation consist of layers of α-$TaN_{(0.1)}$ and β-$TaN_{(0.1)}$ and their in plane dimensions were about 3mm x 7mm. The samples were mounted on a copper sample stage which temperature was controlled by calibrated ruthenium oxide thermometer Pic.13.

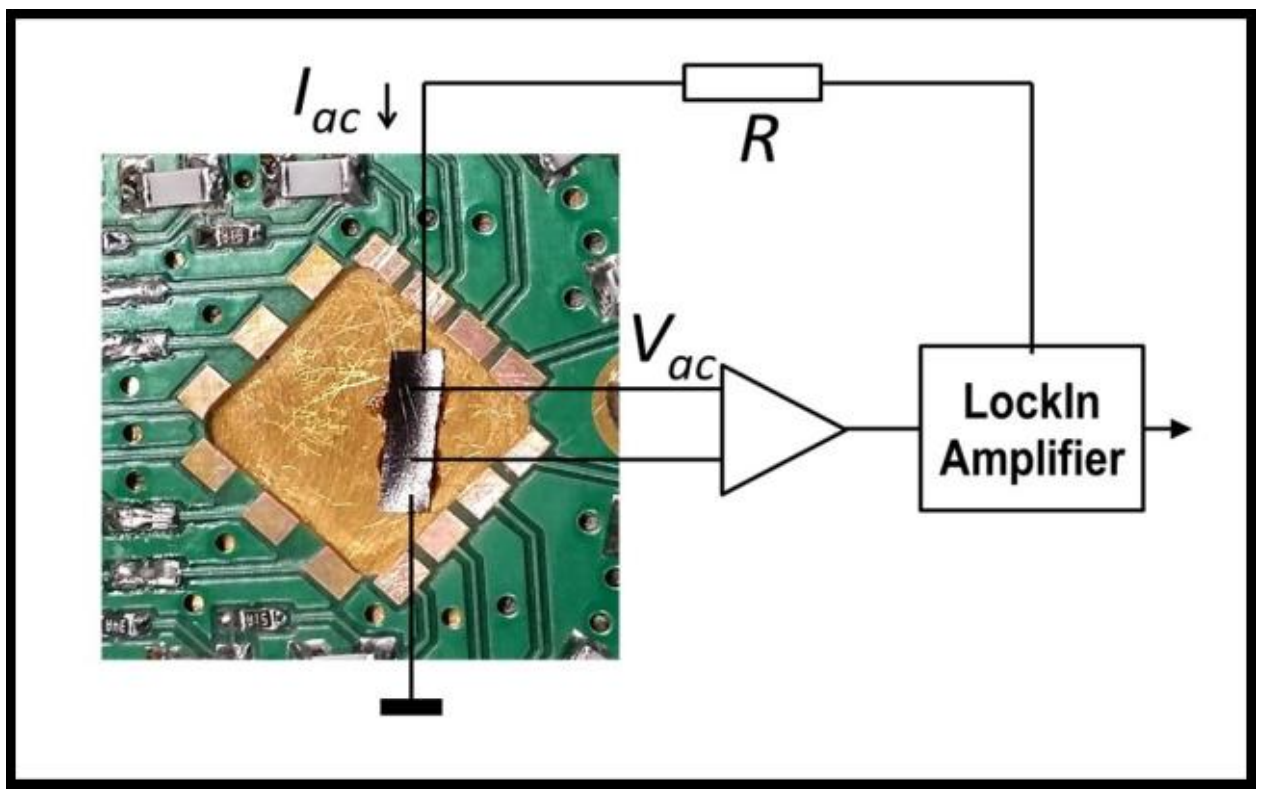

Pic.13. Sample mounted at the cryostat sample holder and schematic diagram of the measurement configuration.

All sample connection leads contain three stage low pass RC filters with cut of frequency 10 kHz and 2 meter thermocoax cable mounted at the sample stage. Electrical resistance of the sample was measured using four probe configuration. In separate measurements all electrical contacts were bonded to one side of the sample (to α-$TaN_{(0.1)}$ or to β-$TaN_{(0.1)}$ layer) using 25μm aluminum wires. Typical resistance of the samples at room temperature were of the order 4 mOhm for samples with contacts bonded to α-$TaN_{(0.1)}$ layer and about 25 mOhm for configuration with contacts to β-$TaN_{(0.1)}$ layer. Lock-in amplifier technique was utilized for the resistance measurements. Investigated samples were biased with a low level AC current (typically 1-10uA, 176 Hz). Voltage across the sample after amplification by low noise voltage preamplifier was measured by SR830 lock-in amplifier connected to computer.

Multilayer structure of the measured samples may significantly affect the distribution of the electrical current in the sample and does not allow make accurate numerical estimations for electrical resistance temperature dependence of separate layers. Nevertheless, significant change of the resistance (e.g., transition in superconducting state) should lead to well pronounced singularity independently on the contact configuration.

The sample resistance vs. bath temperature is presented in Pic.14 for two different samples with electrical contacts bonded to opposite sides of the film (to α-$TaN_{(0.1)}$ or to β-$TaN_{(0.1)}$ layer).

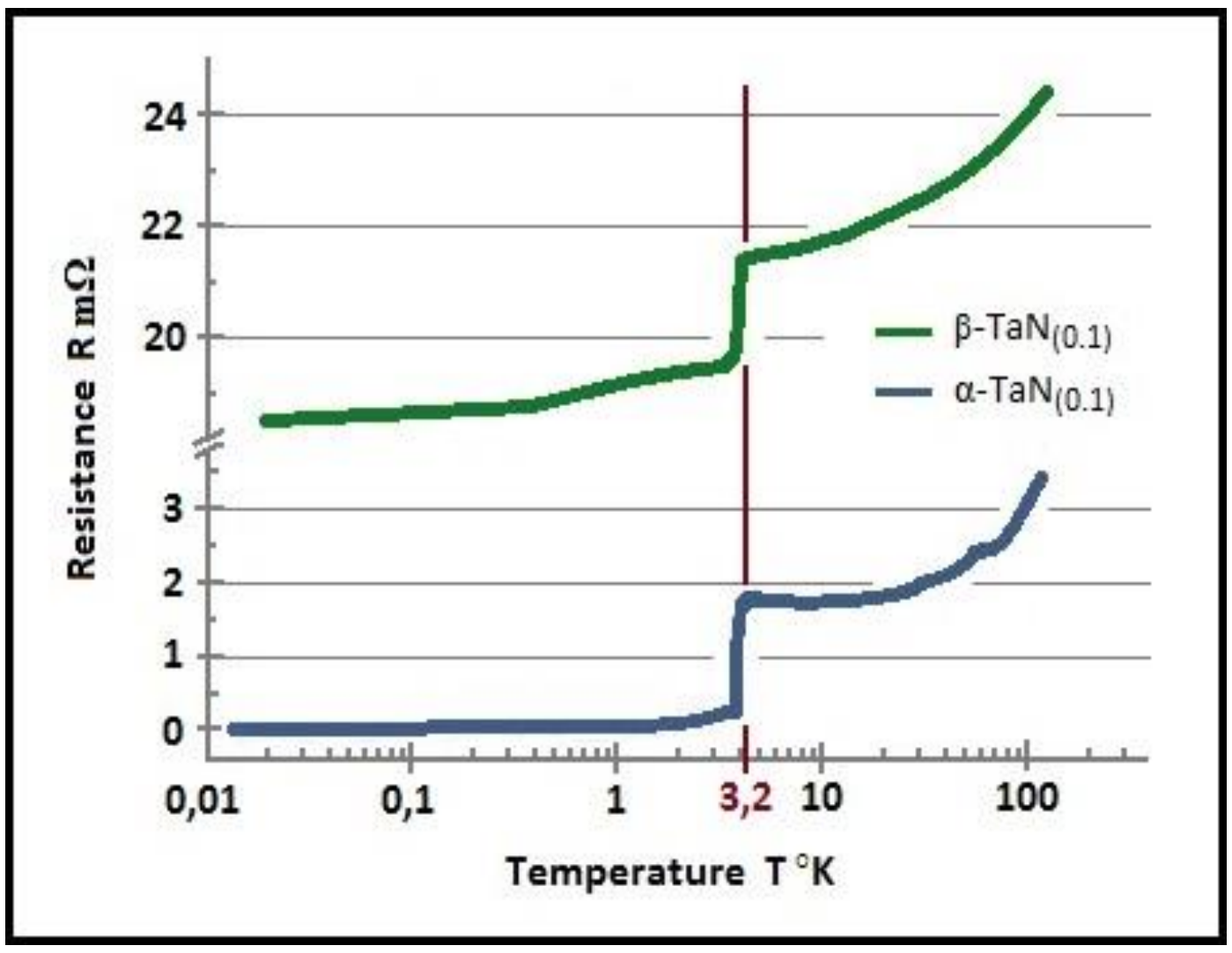

Pic.14. Temperature dependence of electrical resistance for α-TaN(0.1)/β-TaN(0.1) samples, Upper curve corresponds to the sample with contacts attached to β-TaN(0.1) layer, lower curve - sample with contacts attached to α-TaN(0.1) layer.

In both cases sample electrical resistance demonstrates characteristic to metals decrease with decrease of temperature and at bath temperature around 3.2K the resistance was found to exhibit a distinctive step, which we associate with transition to superconducting stage. In the case of the sample with contacts attached to α-$TaN_{(0.1)}$ layer electrical resistance drops practically to zero value at 3.2K and remains at this value at temperatures down to 12mK – the lowest temperature of our experiment. Non zero residual electrical resistance below 3.2K in the sample with contact attached to β-$TaN_{(0.1)}$ layer is probably described by boundary resistance between two layers. Based on the obtained experimental result we conclude that α-$TaN_{(0.1)}$ layer undergoes transition into superconducting state at 3.2K, whereas β-$TaN_{(0.1)}$ remains in normal state at temperatures down to 12mK.

## 4. Summary

The coating materials proved to be of homogeneous structure featuring mainly or close to amorphous phase which however could also be either fine-grained or micro-crystalline. Also, unlike traditional range of PVD coatings, no way the columnar structure of coating materials can happen to be formed by presented technique.

Thicknesses of coatings prepared by the PVD Droplets process are markedly higher than of reference Ta coating (CVD), hence new coatings should serve much longer in various operation conditions and have better corrosion and wear properties.

Hardness of coatings prepared by the PVD Droplets method are much higher, therefore the coatings can be used as real wear and corrosion protective coatings in industrial applications.

The basic coating materials (Ta, Zr, etc.) are known to be highly corrosion resistant materials, thus new applications are available.

The corrosion resistance measurements showed, so these coatings can work in highly corrosive conditions and new applications.

The coatings prepared by the thick coating PVD Droplets process are very dense and then can act as real corrosion barriers.

Obtained results on coatings structuring showed obvious technological advantages of novel PVD Droplets method compared with CVD (Chemical vapor deposition) technology.

Fabricated by suggested method α-$TaN_{(0.1)}$ coating does exhibit superconductive properties while pseudo-cubic crystal structured β-$TaN_{(0.1)}$ coating shows the step in conductivity but, however, not leading to full superconductivity transformation.

The visible defects at the coating-substrate border had been noticed and further were eliminated by correction of technology parameters: concentration of gas plasma was been increased as well the time for ions cleaning and etching of substrate was prolonged in particular.

It has been shown that adjustments in deposition parameters allow to determine material structure variations and therefore to knowingly obtain optimal superficial, mechanical, chemical and other properties of the coating material.

The deposition technique itself is innovative, effective, relatively simple, environment friendly and requires relatively low energy and raw materials supply meaning appears to be costs effective providing high value added production.

As an final summary it is obvious that the coatings should be subject for further development and research to be further realized in industry application and/or sophisticated deposition equipment. This novel coating materials and coatings themselves are definitely to be very interesting from a scientific point of view either.

**Acknowledgments**

*The work has been supported by* TEKES -Finnish Funding Agency for Innovation.

*Authors appreciate the help of staff of* Ota Nano, Low Temperature Laboratory of Aalto University in Critical Temperature testing.

*Our special thanks to* Mr. A. Zhivushkin, *metallurgy professional, for consulting and his contribution in material structures evaluation.*

**References**

[1] Coating Tribology: Properties, Techniques, and Applications in Surface Engineering Holmberg K., Matthews A., (Tribology serious: 28) 1994, ISBN 0444 88870 5

[2] Fundamentals of High Power Impulse. Magnetron Sputtering. Johan Böhlmark. Plasma & Coatings Physics Division. Department of Physics, Chemistry

[3] Corrosion behaviour of magnetron sputtered α- and β-Ta coatings on AISI 4340 steel as a function of coating thickness S. Maeng a,*, L. Axe a, T.A. Tyson b, L. Gladczuk c, M. Sosnowski.

[4] The Structure and Stability of β-Phase Ta Aiqin Jiang *Advisor: Dr. Trevor A. Tyson * Co-Advisor: Dr. Lisa Axe *** Department of Physics** Civil and Environmental Engineering Department New Jersey Institute of TechnologyNewark, NJ

[5] Investigation of the structure of b-Tantalum Aiqin Jiang1, Anto Yohannan1*, Neme O. Nnolim1, Trevor A. Tyson1, Lisa Axe2, Sabrina L. Lee3 and Paul Cote3 1Department of Physics, New Jersey Institute of Technology, Newark, NJ 07102 2Department of Civil and Environmental Engineering, New Jersey Institute of Technology, Newark, NJ 07102 3US Army Armament Research, Development and Engineering Center, Benét Laboratories, Watervliet, NY 12189

[6] R. J. Farraro; Rex B. McLellan. Mechanical Behavior. Received: 22 February 1979. High temperature elastic properties of polycrystalline niobium, tantalum, and vanadium

[7] The Superconductivity of Some Transition Metal Compounds. George F. Hardy and John K. Hulm Phys. Rev. 93, 1004 – Published 1 March 1954.